\documentclass{article}

\PassOptionsToPackage{numbers, compress}{natbib}

\usepackage[preprint]{neurips_2026}

\usepackage[utf8]{inputenc} 
\usepackage[T1]{fontenc}    
\usepackage{hyperref}       
\usepackage{url}            
\usepackage{booktabs}       
\usepackage{amsfonts}       
\usepackage{nicefrac}       
\usepackage{microtype}      
\usepackage{xcolor}         
\usepackage{graphicx}       
\usepackage{subcaption}     
\usepackage{multirow}       
\usepackage{enumitem}       
\usepackage{wrapfig}        
\usepackage{amsmath}

\usepackage{fontawesome}
\usepackage{float}
\usepackage{placeins}
\usepackage{longtable}
\usepackage{xspace}
\usepackage[capitalize]{cleveref}
\usepackage{todonotes}
\crefname{figure}{Fig.}{Figs.}
\Crefname{figure}{Fig.}{Figs.}
\crefname{section}{Sec.}{Secs.}
\Crefname{section}{Sec.}{Secs.}
\newcommand{\method}{\textsc{Neurrator}\xspace}

\title{Can neurons speak? \\Semantic narration of vision at single-cell resolution}

\author{%
  Arnau Marin-Llobet\\
  Harvard University\\
  \And
  Richard Hakim\\
  Kempner Institute\\
  Harvard University
  \And
  Sara Matias\\
  Center for Brain Science\\
  Harvard University\\
  \And
  Venkatesh N. Murthy\\
  Center for Brain Science\\
  Kempner Institute\\
  Harvard University\\
  \And
    Na Li\\
  Harvard University\\
  \And
  Demba Ba \\
  Kempner Institute\\
  Harvard University
}

\begin{document}

\maketitle

\begin{abstract} Identifying what individual neurons encode in higher-order visual cortex is an open problem. Responses resist intuitive parameterization, and the deep-network embeddings used in their place are black boxes. Here, we introduce \method, a framework that decodes spiking activity into free-form natural-language narration of the viewed scene at single-neuron resolution. A learned encoder maps spike trains from arbitrary subsets of simultaneously-recorded neurons into the patch-embedding space of a frozen CLIP, from which a multimodal language model and sparse autoencoder generates and validates a description with no language-side training. Applied to Neuropixel recordings of mouse visual cortex during natural-movie viewing, \method narrates from thousands of neurons, singular cortical regions, local populations, or from a molecularly-defined cell-types. We use this property to (i) quantify how decoding fidelity scales with population size and cortical region, and (ii) \emph{neurrate}, in plain language, what individual neurons and genetically-tagged inhibitory cell-types contribute to visual representation. This recasts cell identity from a classification target into a functional probe of the visual system, providing a new unit of biological insights in neural systems. \\Code available at \faGithub: \url{https://github.com/arnaumarin/neurrator}
\end{abstract}

\section{Introduction}
A central problem in neuroscience is identifying and explaining what individual neurons encode. The most common strategy is to parameterize an external variable and associate those parameters with neural activity patterns. For example, many retinal ganglion cells respond to spots of light or darkness, which can be efficiently parameterized by the spot's position, size, and polarity \cite{kuffler1952neurons, stanley1999reconstruction}. This approach breaks down in higher-order visual cortical areas, where neurons respond to complex visual features that are not easily parameterized along intuitive axes \cite{maunsell1983connections, felleman1991distributed, gross1973inferotemporal, kobatake1994neuronal, tsao2008comparing, vinken2023neural}. To extend tuning analysis into this regime, recent work ``parameterizes'' natural images and videos by embedding them into the latent representational spaces of large neural networks, then maps these latents onto neural activity \cite{yamins2014performance, yamins2016using, khaligh2014deep, schrimpf2018brain, schrimpf2021neural, bashivan2019neural, ponce2019evolving, cadena2019deep, kell2018task, cunningham2014dimensionality, kriegeskorte2019interpreting, schneider2023learnable}. This substantially improves predictive performance over hand-designed feature spaces, but the investigator must still translate high-dimensional activations into semantic hypotheses, and ultimately a plain-language description, using manual stimulus inspection, retrieval, or attribution analysis.

The tension between modeling for raw predictive power vs. interpretability defines a Pareto front for decoding neural activity: the most predictive targets are often black-boxes. Contrastive vision-language models offer a way past this bottleneck. Models such as CLIP, ALIGN, and SigLIP learn a joint embedding in which images and their natural-language descriptions occupy nearby points \cite{radford2021clip, jia2021scaling, zhai2023sigmoid}, and multimodal language models built on top of these encoders (BLIP-2, Flamingo, LLaVA) take embeddings from this space as inputs and emit free-form natural-language descriptions \cite{li2023blip, alayrac2022flamingo, liu2023llava}. The combination of these two types of models offers a bridge to language for any signal that can be associated with the embedding space of the vision-language model. The same space also admits a feature-level decomposition: sparse autoencoders (SAEs) fit to its activations expose a finite dictionary of interpretable visual-concept directions \cite{josephsteering2025, bricken2023monosemanticity, fel2025archetypal}. Together, these two properties (generative read-out and concept-level decomposition) make this space a natural target for neural decoding: a model that maps spikes into it would produce, for free, both human-readable descriptions of what a population represents and a principled basis for asking which visual concepts each neuron contributes to. To our knowledge, no prior work has used this space as the target of a single-unit decoder: existing electrophysiological decoders either reconstruct low-level stimulus features or remain confined to opaque embedding coordinates, leaving the description bottleneck in place. Closing it would do more than improve readability of decoded outputs --- it would open a new mode of inquiry in which hypotheses about neural systems can be posed, queried, and tested directly in natural language, at the resolution of individual and interest-specific populations.

\begin{figure}[h]
    \centering
    \includegraphics[width=\linewidth]{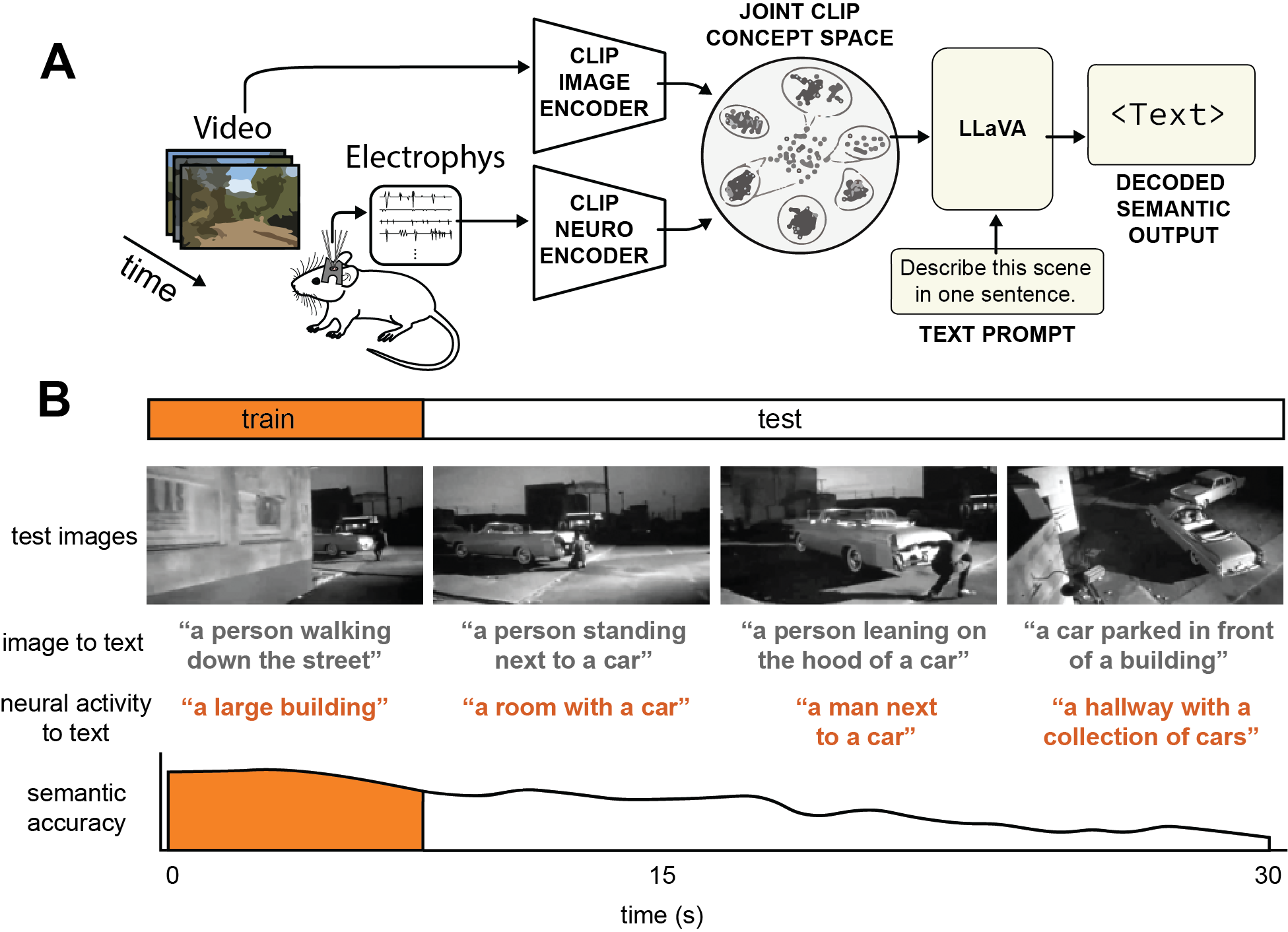}
    \caption{\textbf{\method: language-aligned readout from spiking activity.}
    \textbf{(A)}~A learned neural encoder maps spike trains into the joint CLIP embedding space shared with the frozen CLIP image encoder; a frozen LLaVA then decodes the predicted embedding into a free-form description of the viewed scene.
    \textbf{(B)}~Representative example decodings on held-out test frames from a natural movie. Image-to-text captions (gray) are produced from the video; neural-activity-to-text captions (orange) are produced from spiking activity alone. Bottom: semantic accuracy over the clip.}
    \label{fig:video_narration_main}
\end{figure}

We instantiate this idea as \method, a framework that maps the spiking activity of individual neurons directly to natural-language narration of the viewed visual scene. \method consists of a learned encoder that takes the spike trains of a chosen subset of recorded neurons and predicts the corresponding patch embeddings of a frozen vision tower; a frozen multimodal language model then decodes those embeddings into a free-form description (Fig.~\ref{fig:video_narration_main}; additional examples in Figs.~\ref{fig:video_narration}, \ref{fig:video_narration_nm3}). We train and evaluate \method on Neuropixels recordings of mouse visual cortex during natural-movie viewing \cite{siegle2021allen, steinmetz2021neuropixels}, with CLIP ViT as the target embedding space and LLaVA as the language decoder \cite{radford2021clip,liu2023llava}. Because the encoder is uniform over input subsets, the same trained model can be queried on arbitrary subpopulations: we use this to quantify how semantic accuracy scales with the number of input neurons, the cortical region they are drawn from, and the cell-type composition of the population. To move beyond raw text and recover a structured account of what each subpopulation encodes, we then push the predicted embeddings through the pretrained CLIP space by fitting SAE \cite{josephsteering2025} to create dictionaries of principled visual-concept features. Empirically, we find distinct concept signatures across different brain regions and  genetically defined cell types.

Our contributions are as follows:
\begin{itemize}[leftmargin=*]
    \item \textbf{Semantic-based neural decoder.} We introduce \method, the first decoder that maps single-spike-level population activity directly to semantically coherent natural-language descriptions of visual experience, generalising across held-out frames, held-out image identities, and an unseen second movie (Sec. \ref{sec:narration}).
    \item \textbf{Region- and cell-type identity as a functional probe.} Because the decoder is uniform over input subsets, we restrict it at inference to single neurons, anatomical regions, or molecularly-defined cell types and read out, in language, what each subset contributes to the representation. This yields scaling laws for semantic decoding fidelity as a function of population size and cortical region, and recasts cell-type and region labels from classification targets into functional probes of visual processing (Sec. \ref{sec:scaling}).
    \item \textbf{Concept-level decomposition of cell-type contributions.} Combining the above with a pretrained CLIP sparse autoencoder \cite{josephsteering2025}, we decompose each subpopulation's contribution into interpretable visual-concept features, recovering cell-type-distinct concept signatures that serve as hypothesis (e.g.\ PV $\to$ small rounded objects) under bootstrap resampling and an orthogonal CLIP-text concept-axis validation (Sec. \ref{sec:sae_dict}).
\end{itemize}

\section{Related work}
\label{sec:related_work}
\paragraph{Generative language readouts from human neural data.}
The shared vision--language space has been used extensively as both encoding target and decoding source for non-invasive human recordings. Vision-language and language-model embeddings of natural stimuli predict fMRI BOLD responses \cite{liu2023brainclip, conwell2024large, doerig2025high, huth2016natural, caucheteux2023evidence, tang2023brain}, fMRI decoders trained against image space reconstruct viewed images \cite{takagi2023high, scotti2024mindeye2, halac2022multiscale}, and decoders trained against language-model representations reconstruct continuous natural language from perceived speech, imagined speech, and silent video \cite{tang2023semantic, defossez2022decoding}. Recent work has even captioned the preferred stimulus of individual voxels in free-form natural language \cite{luo2024brainscuba}, taking a step toward per-unit interpretability. The fundamental limitation is spatial: each voxel or electrode contact integrates over $10^4$--$10^6$ neurons, so even per-voxel readouts describe a region rather than a cell. \method shares this per-unit ambition but operates three orders of magnitude finer, on a substrate where molecular cell-type identity is independently recoverable, and yields a per-trial trajectory rather than a single tuning summary.

\paragraph{Sparse autoencoders as probes of vision--language space.}
Sparse autoencoders (SAEs) decompose dense embedding activations into a dictionary of sparsely-activating, interpretable feature directions \cite{josephsteering2025, bricken2023monosemanticity}, converting vector-valued activations into a sparse profile over named concepts. Currently, SAEs are the most effective tool we have for interpreting learned representations \cite{josephsteering2025, simon2025interplm, gujral2025sparse}. To date, however, this technology has been turned almost exclusively inward, applied to the activations of foundation models themselves rather than used as a probe into the neural systems those models are intended to illuminate. The few applications of SAE and related methods for interpretability of neural data have so far operated on indirect, population-averaged signals such as calcium imaging or local field potentials \cite{freeman2025beyond, marin2025neural}, which integrate over many cells and lack the temporal precision of spiking activity. \method makes the connection at the level of single-unit spike trains: because spikes are projected into the same shared vision--language space on which SAEs operate, biological population activity can be read out simultaneously as a free-form sentence and as a sparse profile over named visual concepts, both produced from the identical neural-side embedding.

\paragraph{Cell-type identity as input rather than output.}
A growing body of work treats \textit{in vivo} cell-type and brain-region identity as the \emph{output} of a classifier trained on extracellular features, using unsupervised multi-modal embeddings of waveform shape and spike-train statistics \cite{physmap}, supervised classifiers calibrated against optogenetic ground truth \cite{beau2025}, multi-modal contrastive pretraining \cite{yu2025vivo}, or general-purpose vision-language models repurposed as few-shot subtype classifiers \cite{bciagent}. In each case the identity label is the endpoint of the analysis. \method takes the complementary view: cell-type and brain-region identity (whether obtained from optotagging or any of the methods above) enter the model as \emph{inputs}, and the model returns a free-form description of what activity from that subpopulation is encoding on a given trial. \method thus recasts cell-type identity from a classification target into a functional probe of the neural system.
\section{Approach} 
\label{sec:approach}
\subsection{\method framework}

\begin{figure}[ht]
    \centering
    \includegraphics[width=\linewidth]{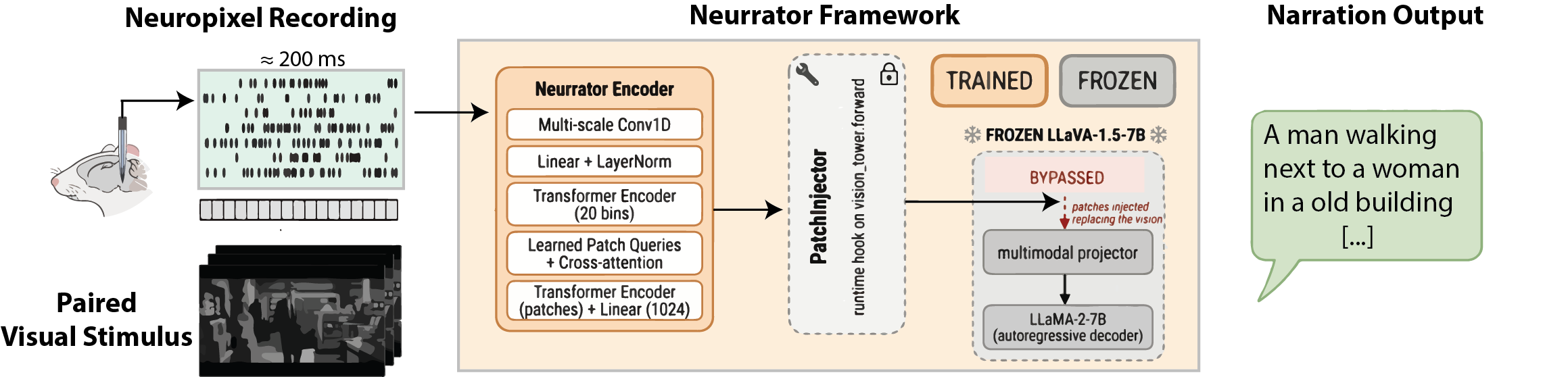}
    \caption{\textbf{Overview of the Neurrator framework}. A trainable Neurrator Encoder processes Neuropixel recordings from a mouse viewing a visual stimulus, mapping spike trains to visual patch embeddings via multi-scale Conv1D layers, transformer encoders, and learned patch queries with cross-attention. A PatchInjector hooks into the frozen LLaVA model at runtime, replacing the output of its vision tower with the predicted patches. The frozen multimodal projector and LLaMA decoder then generate a natural language narration of the perceived stimulus.}
    \label{fig:neurrator_architecture}
\end{figure}

\method maps spike trains recorded with high-density Neuropixels probes \cite{steinmetz2021neuropixels} to natural-language descriptions of the visual stimulus the animal is viewing, by routing neural activity through the embedding space of a vision--language model. Spike counts from all single units passing Allen-Institute quality control are binned and $z$-scored per neuron using statistics from training repeats only, and a short window of activity is fed to the trainable {\method Encoder}(Fig.~\ref{fig:neurrator_architecture}). The encoder's output is a patch-embedding tensor with the exact shape that CLIP ViT-L/14 \cite{radford2021clip} produces at its penultimate layer for a real movie frame: $576$ patch tokens (a $24\!\times\!24$ grid) of dimension $1024$. This patch tensor is the only learned interface between brain and language: it is handed verbatim to a frozen LLaVA-1.5-7B \cite{liu2023llava}, whose vision tower is bypassed at runtime by a forward hook ({PatchInjector}; Fig.~\ref{fig:neurrator_architecture}). The multimodal projector and the LLaMA-2-7B decoder \cite{touvron2023llama} operate as in standard image captioning, treating the neurally-derived patches as if they had been produced by the actual image. No part of the language model is ever trained on neural data. The encoder itself uses a multi-scale 1-D-convolutional spike-train front end, a small transformer over the temporal window, attention-weighted temporal pooling, and $576$ learned patch queries that cross-attend to the pooled representation to produce one $1024$-d embedding per CLIP patch; full details are in Appendix~\ref{app:method_details}.

\paragraph{Training objective}
Targets are obtained by running CLIP ViT-L/14 (the exact vision tower inside LLaVA-1.5-7B) on every stimulus frame and extracting the penultimate-layer hidden states; the resulting $(N_\text{frames}, 576, 1024)$ tensor is the supervision signal. The encoder is trained to regress its prediction $\hat{P}_t$ onto the true patch tensor $P_{f(t)}$ for the frame $f(t)$ presented at time $t$, under a dual loss combining mean-squared error and per-patch cosine similarity in equal proportion, \[ \mathcal{L} \;=\; 0.5\cdot \mathrm{MSE}(\hat{P}_t, P_{f(t)}) \;+\; 0.5\cdot \bigl(1 - \cos(\hat{P}_t, P_{f(t)})\bigr), \] which keeps both the magnitudes and the directions of the predicted patches aligned with what the frozen LLaVA expects. Optimizer, schedule, and training-budget details are in Appendix~\ref{app:method_details}.

\paragraph{Natural language decoding}
At test time we never run LLaVA's CLIP encoder. Instead, the encoder prediction $\hat{P}_t$ is reshaped and returned as the penultimate-layer hidden states of LLaVA's vision tower through a forward hook. Everything downstream of the vision tower (the multimodal projector and the LLaMA-2-7B language decoder) is left untouched and runs in its default greedy-decoding configuration with a fixed one-sentence-description prompt (Appendix~\ref{app:method_details}). Because the patch tensor is the only modality bridge, every difference between cell types, brain regions, or training conditions reduces to a difference in those $576\!\times\!1024$ predicted features; the language model itself sees no neural data and contributes only its image-conditional prior.

\subsection{Datasets} 
\label{methods:datasets}
We use $16$ recording sessions from the Allen Brain Observatory Visual Coding Neuropixels release \cite{siegle2021allen} (Brain Observatory 1.1 subset, the only protocol that includes naturalistic stimuli). Each mouse views three classes of natural content during the same recording: \emph{Natural Movie One} (NM1; {30}{s} clip, $900$ frames at {30}{Hz}, $20$ repeats), \emph{Natural Movie Three} (NM3; {120}{s}, $3{,}600$ frames, $10$ repeats), and \emph{Natural Scenes} ($118$ grayscale photographs, $\sim\!50$ presentations each). For cell-type analyses we intersect with the Siegle et al.\ optotagging tables, yielding $73$ PV, $49$ SST, and $33$ VIP optotagged neurons across the cohort; frame-aligned pseudo-mice are constructed by concatenating optotagged columns across sessions of matching genotype. Full dataset, optotagging-criteria, and pseudo-mouse-construction details are in Appendix~\ref{app:datasets}.

\section{Results}
\label{sec:results}

\subsection{Spikes-to-sentences are semantically coherent and generalize to held-out frames}
\label{sec:narration}

\begin{figure}[h]
  \centering
  \includegraphics[width=\textwidth]{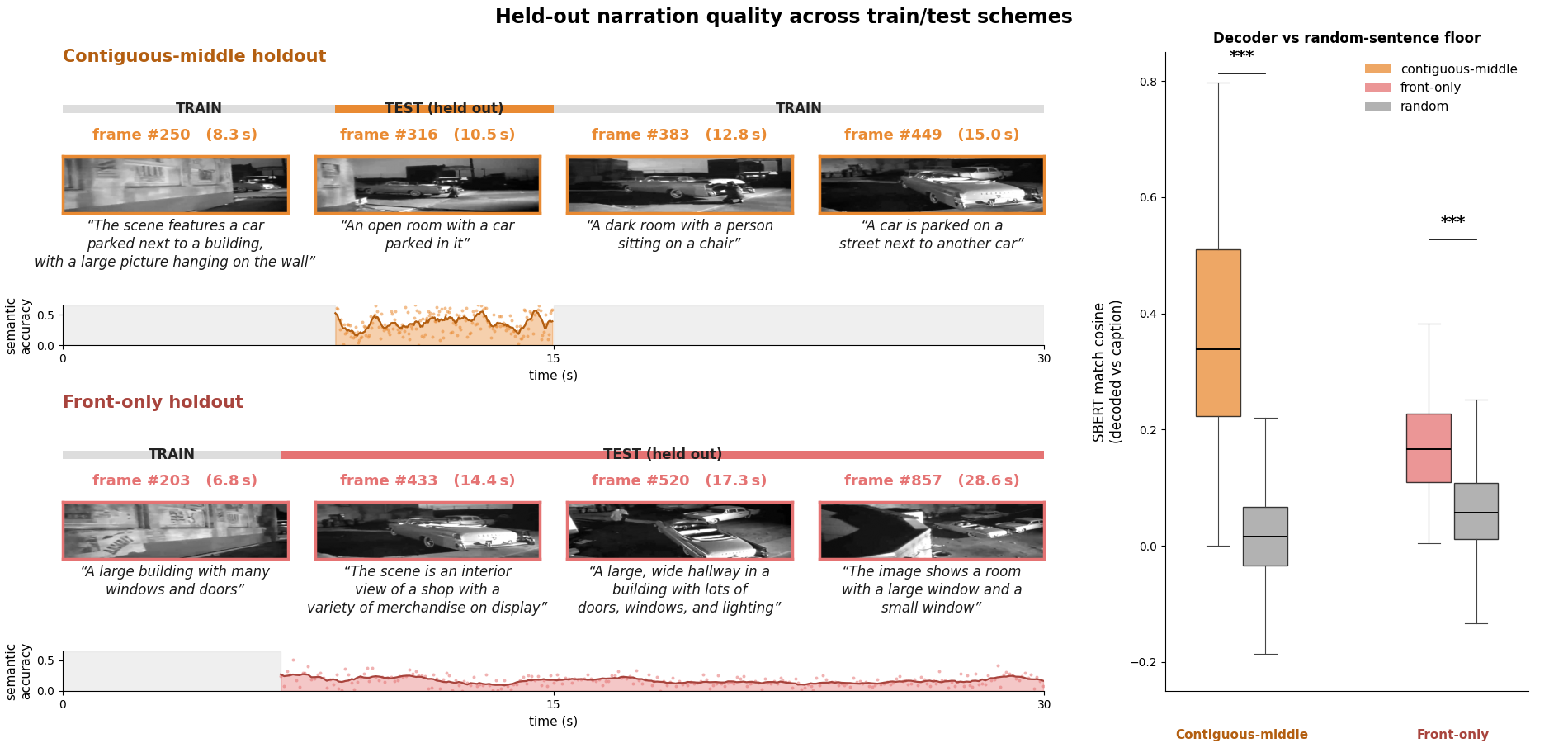}
  \caption{\textbf{Held-out narration quality.} Left: semantic accuracy over time for contiguous-middle and front-only holdouts. Right: SBERT cosine vs random-sentence floor; *** $p<0.001$.}
  \label{fig:held_out_narration_quality}
\end{figure}

\paragraph{Natural-language semantic decoding of natural visual stimuli} \method produces content-accurate natural-language narrations of natural movies from single-neuron spike-trains alone (Fig.~\ref{fig:held_out_narration_quality}). To establish that these narrations reflect the visual stream rather than memorisation of training frames, we evaluate two stress-test partitions in which entire blocks of the movie are excluded from language decoder training. In the \emph{contiguous-middle} regime, frames 250--449 (a continuous scene of $\sim$6.7\,s) are held out, forcing the decoder to interpolate over an unseen scene from temporally distant training context; in the \emph{front-only} regime, training is restricted to the first 200 frames and the remaining 700 frames must be extrapolated.

Across both regimes, decoded narrations on held-out frames remain semantically aligned with the visual content. We use Sentence-BERT (SBERT) as a metric to measure its semantic similarity (SBERT cosine: a sentence-level semantic similarity score in $[-1, 1]$, where $\sim 1$ means the two sentences are semantically the equal and $\sim 0$ means they are unrelated) \cite{Reimers2019SentenceBERTSE}. On the contiguous-middle test block, mean SBERT cosine between decoded and BLIP-2 reference captions is $0.367 \pm 0.180$ ($n=200$ frames) versus $0.020 \pm 0.077$ for a word-salad floor ($\Delta = +0.347$, $p<0.001$). The front-only regime, which requires extrapolation across $>$3$\times$ the training horizon, still yields $0.170 \pm 0.085$ versus a $0.062 \pm 0.073$ random floor ($\Delta = +0.108$, $p<0.001$). Inspection of the example narrations (Fig.~\ref{fig:held_out_narration_quality}, thumbnails) confirms that decoded sentences correctly recover scene-level structure on never-seen frames---car-parking layouts (frame~316: ``an open room with a car parked in it''), interior scenes (frame~383: ``a dark room with a person sitting in a chair''), and architectural detail in the extrapolation regime (frame~857: ``the image shows a room with a large window and a small window''). The semantic-accuracy curves further show that decoding quality is highest near the train/test boundaries and degrades smoothly with distance from training context.

Replicating the full pipeline on a different movie (NM3) yields narrations that describe the distinct content of that film (people, suits, bicycles, groups), with no re-training of the language decoder and no stimulus-specific prior (Fig.~\ref{fig:video_narration_nm3}). To our knowledge, no prior work has produced free-form natural-language descriptions of the visual stream from single-unit electrophysiology, nor demonstrated that such descriptions generalise across held-out scenes.

\paragraph{Generalization to never-seen image identities.} Held-out frames within a familiar movie still share statistics with their training neighbors. A stronger test is whether \method can decode neural responses to \emph{novel image identities} that the encoder has never been exposed to. Using the Allen Natural Scenes stimulus, we hold out $18$ of $118$ grayscale photographs at the identity level (their spike responses appear only at test time) and decode narrations from the held-out trials. Decoded sentences score significantly higher against the true BLIP-2 caption than against a shuffled-pairing control (matched SBERT $0.282 \pm 0.178$ vs.\ shuffled $0.222 \pm 0.136$, $\Delta\!=\!+0.060$, $p\!<\!10^{-2}$), with content-accurate decodes on unseen images --- e.g.\ a tiger-in-tall-grass scene narrated as \textit{``a tiger walking through the grass''} ($+0.68$). Full numerics, controls, and examples are reported in Appendix~\ref{app:image_holdout}.
\begin{figure*}[t]
    \centering
    \includegraphics[width=\linewidth]{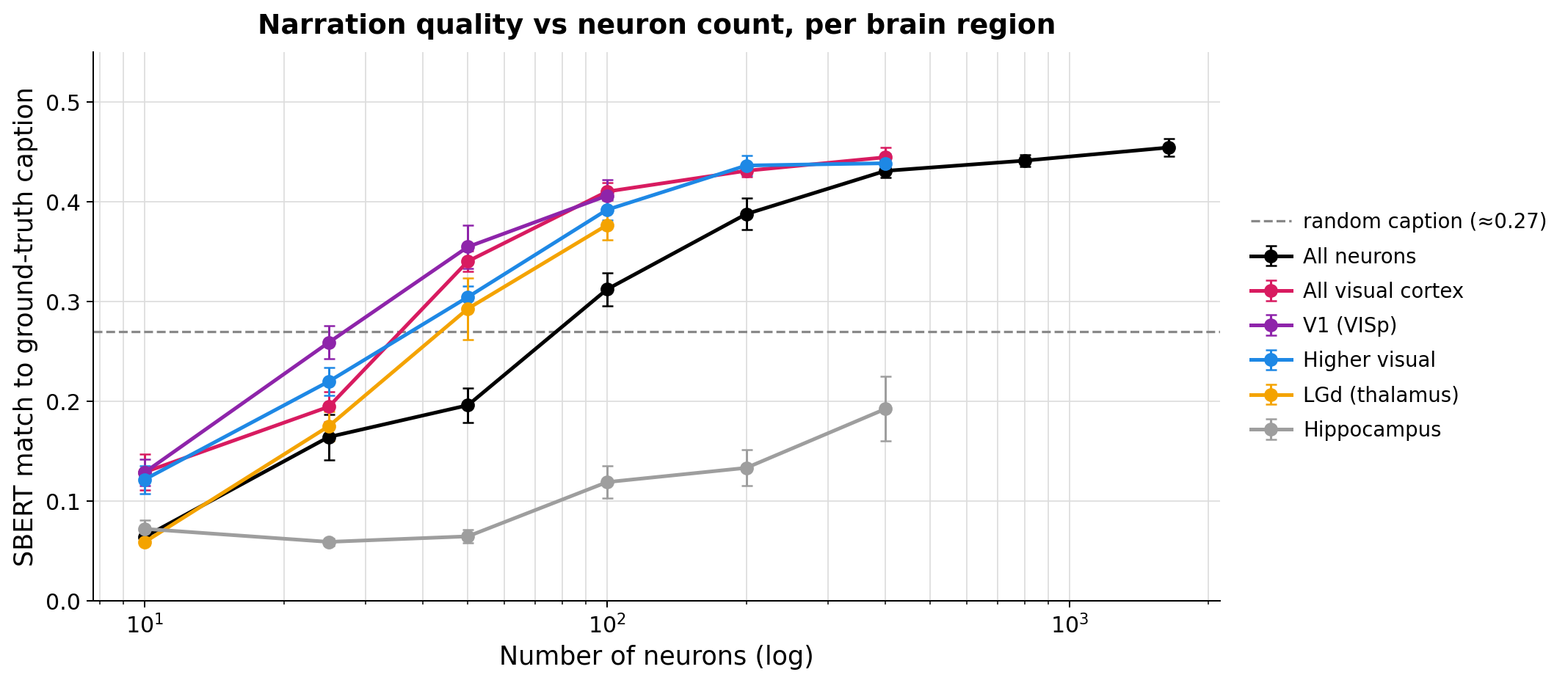}
    \caption{\textbf{Decoding scales with neuron count across visual areas and animals.}
    SBERT similarity between decoded narrations and ground-truth captions vs.\ number of neurons, by region.} 
    \label{fig:scaling_by_area}
\end{figure*}
\subsection{Scaling of semantic decoding across visual areas}
\label{sec:scaling}
\paragraph{Narration fidelity scales with population size and requires ${\sim}10^2$ visually-driven neurons.}
Figure~\ref{fig:scaling_by_area} plots held-out SBERT similarity between decoded narrations and ground-truth captions against input population size, broken down by anatomical pool. Across all visual regions (V1, higher visual cortex, LGd, and the union of visual cortex), narration quality scales monotonically on a log axis, rising from near-random levels at $\sim$10 neurons and continuing to climb across the full range tested, reaching $\sim$0.45 SBERT cosine for the largest populations without showing signs of saturation. The dashed line at SBERT~$\approx$~0.28 marks the score of a random caption against ground truth---the level at which decoded sentences carry no more scene-specific content than any generic English sentence would by chance. Visual pools cross this floor only once tens to ${\sim}10^2$ neurons enter the encoder: V1 is the most efficient (${\sim}30$ neurons), higher visual cortex and LGd cross at $50$--$100$, and the heterogeneous all-neurons pool only at ${\sim}100$. Below this range, decoded narrations sit at or under the random-caption baseline and the readout is not yet semantically grounded. Hippocampus, included as a non-visual control, never crosses the random-caption line across the full range tested, consistent with little stimulus-locked visual content under passive natural-movie viewing. Notably, the all-neurons pool lags the visual-only pools at matched population size, indicating that the bottleneck on narration fidelity is the count of \emph{visually-driven} neurons rather than raw spike count.

\begin{figure*}[h] \centering \includegraphics[width=0.93\linewidth]{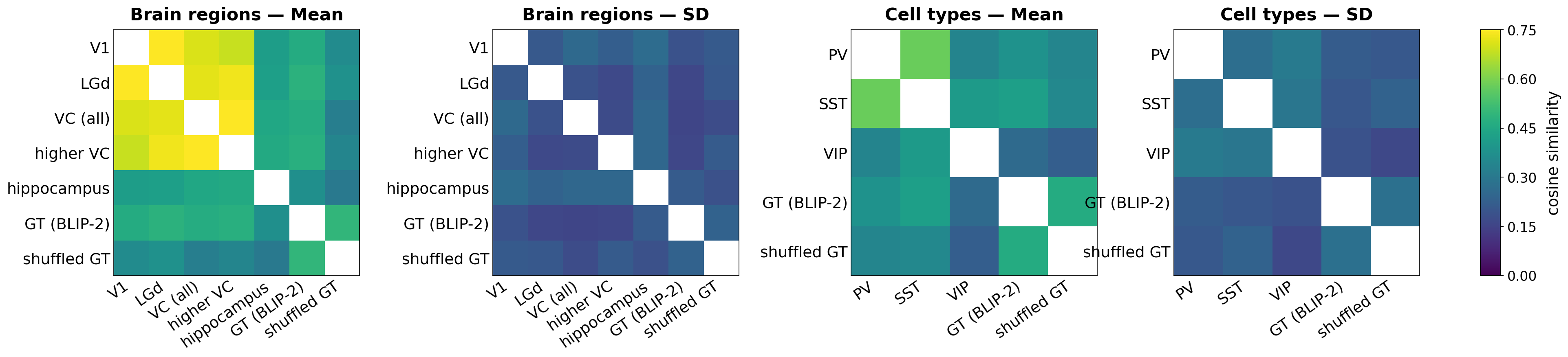} \caption{\textbf{Cross-narration SBERT similarity by brain region and cell type} (NM1 test frames, shared cosine scale 0--0.75). Region pools (left) collapse onto a single visual cluster; only hippocampus drops near the shuffled-GT floor. Cell-type pools (right) show PV--SST clustering with VIP separated.} \label{fig:sim_matrices_combined} \end{figure*}

\paragraph{Region pooling collapses; cell-type pooling separates.}
We next asked whether different subpopulations produce semantically \emph{distinct} narrations of the same stimulus. Pooling neurons either by anatomical region (V1, higher visual cortex, LGd, hippocampus, all visual cortex) or by genetically-tagged cell (PV, SST, VIP), we computed pairwise SBERT cosine between the narration sets and observed two patterns (Fig.~\ref{fig:sim_matrices_combined}). At the regional level, all visual areas collapse onto a single semantic cluster ($0.69$--$0.79$ pairwise cosine), with only hippocampus dropping near the shuffled-pair floor---anatomy alone is a weak handle on narration content. At the cell-type level the picture inverts: PV and SST narrations are modestly aligned ($0.58$), but VIP separates from both ($0.34$ vs PV, $0.41$ vs SST). This narration-level cell-type divergence is the entry point for our discovery-tool application (Sec.~\ref{sec:application}).
\section{Application: cell-type-specific semantic interrogation}
\label{sec:application}
The cell-type divergence in Fig.~\ref{fig:sim_matrices_combined} suggests that genetically defined populations carry semantically distinct readouts of the same stimulus, but it does not tell us \emph{what} each population is encoding. We further use \method's subset-query property to test this: because the encoder is uniform over input neurons, the same trained model can be restricted at inference to a chosen population and asked, in language, what the visual world looks like through the lens of just those cells. We apply this to the three optotagged inhibitory populations from \cite{siegle2021allen} (PV, SST, and VIP interneurons, details in Sec.~\ref{methods:datasets}) and ask whether their narration differences correspond to recognisable visual concepts. Because each Cre line yields only a few tens of optotagged units per session, we construct a frame-aligned ``pseudo-mouse'' per genotype by concatenating the optotagged columns across sessions of the same line, possible because every animal views the identical NM1 frame sequence. The same trained \method is then queried once per pseudo-population.

\begin{figure*}[h]
    \centering
    \includegraphics[width=0.75\linewidth]{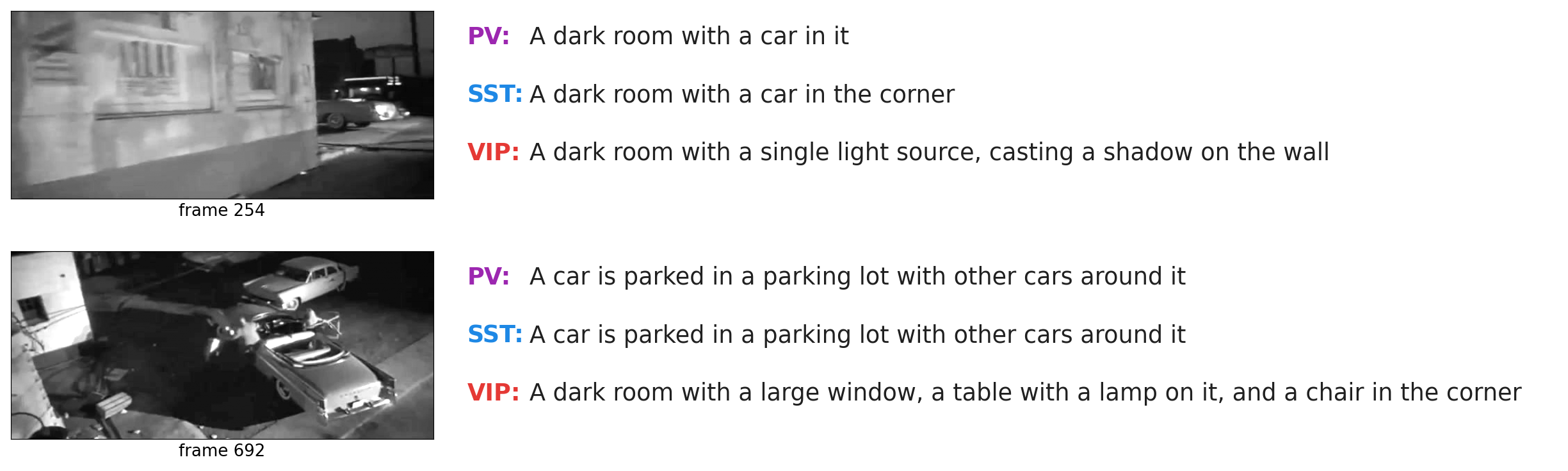}
    \caption{\textbf{Per-frame narration excerpts on NM1.}
    Two test frames where PV and SST describe the visible cars while
    VIP foregrounds lighting and shadow of the same scene.}
    \label{fig:celltype_narration_examples}
\end{figure*}

\begin{figure*}[h]
    \centering
    \includegraphics[width=0.85\linewidth]{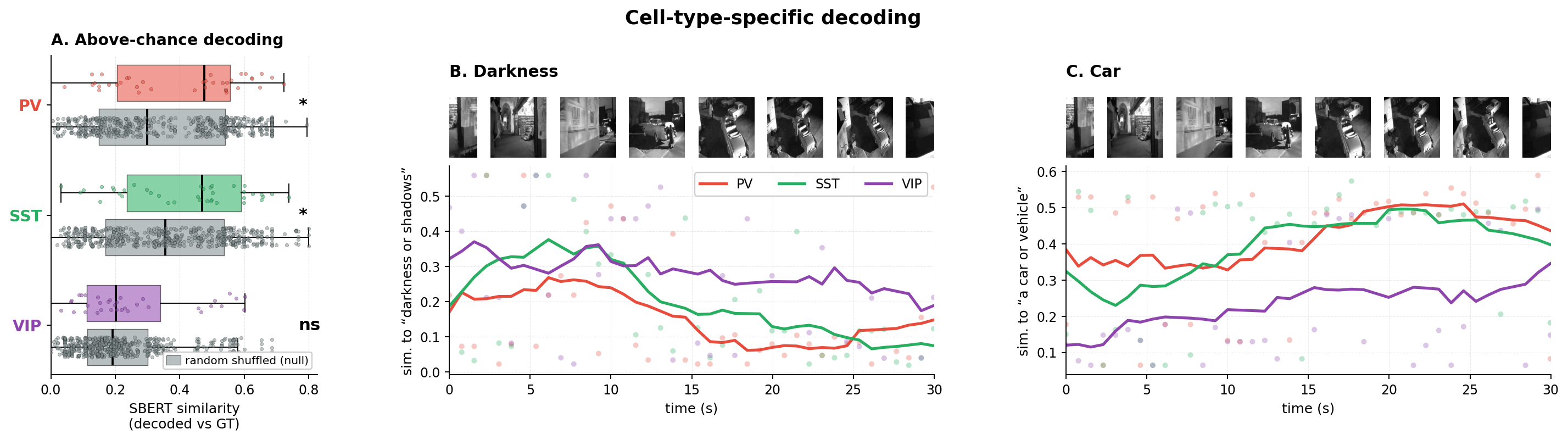}
    \caption{\textbf{Cell-type-specific decoding on NM1.} (A) Per-frame SBERT similarity to BLIP-2 captions. (B,C) Time-resolved cosine of decoded narrations to \textit{``darkness or shadows''} (B) and \textit{``a car or vehicle''} (C); lines: smoothed per-cell-type means, dots: individual frames.}
    \label{fig:celltype_main}
\end{figure*}

\paragraph{Optotagged cell types produce semantically distinct narrations.} Switching the population label from anatomy to genetic identity inverts the picture seen at the regional level. To compare what the three cell-type pools say about the same movie, we compute the SBERT cosine between their decoded narrations on NM1. PV and SST narrations remain modestly aligned ($0.58$ on average), while VIP sits apart from both (Fig.~\ref{fig:sim_matrices_combined}, right). The same gap shows up at the level of individual word use: on Natural Scenes, the most distinctive words (highest log-odds against the other two populations) are \textit{tree / foreground / background} for PV, \textit{building / boat / water} for SST, and \textit{scene / lot / bushes} for VIP, and a simple 3-way classifier trained on the narration embeddings can identify the source cell-type with $76\%$ accuracy (chance $33\%$, $p<10^{-4}$). On NM1, projecting decoded sentences onto the \textit{``darkness or shadows''} and \textit{``a car or vehicle''} axes (i.e.\ measuring how strongly each decoded narration mentions these two concepts over time) shows that all three populations track the gross content of the movie but with different baselines and amplitudes (Fig.~\ref{fig:celltype_main}). The gap is most visible at the single-sentence level (Fig.~\ref{fig:celltype_narration_examples}): on the same NM1 frame, PV and SST routinely produce content-accurate \textit{``a car is parked in a parking lot, and the driver is getting out of the vehicle''}-style descriptions, while VIP describes the same scene as \textit{``a dark room with a single light source, casting a shadow on the wall''} --- a lighting-and-atmosphere reading of the same visual input.

\paragraph{What does each cell-type actually \emph{see}?} \label{sec:sae_dict} Narrations hint at differences but reveal nothing about the underlying visual content driving them. To recover recognisable visual concepts, we turn to SAEs, currently the most effective tool we have for interpreting learned representations~\citep{bricken2023monosemanticity}: a small unsupervised model decomposes a high-dimensional embedding into a much larger dictionary of features, only a handful of which activate per input, so each can be inspected individually. We use the Prisma-Multimodal SAE pretrained on the CLIP\,B/32 layer-11 residual stream (access to $49{,}152$ features) and pass each cell-type's \method patches predictions through it. For each cell-type we rank features by mean activation across the NM1 test bins and keep its top-$20$; features that appear uniquely in a single cell-type's top-$20$ we call \emph{unique-by-magnitude}. To label these without leaking information from the narrations themselves, we run the SAE on a held-out corpus ($50{,}000$-image ImageNet-1k set \cite{russakovsky2015imagenet}) and assign each feature the visual concept unifying its top-$32$ activating images.

\begin{figure*}[h]
    \centering
    \includegraphics[width=\linewidth]{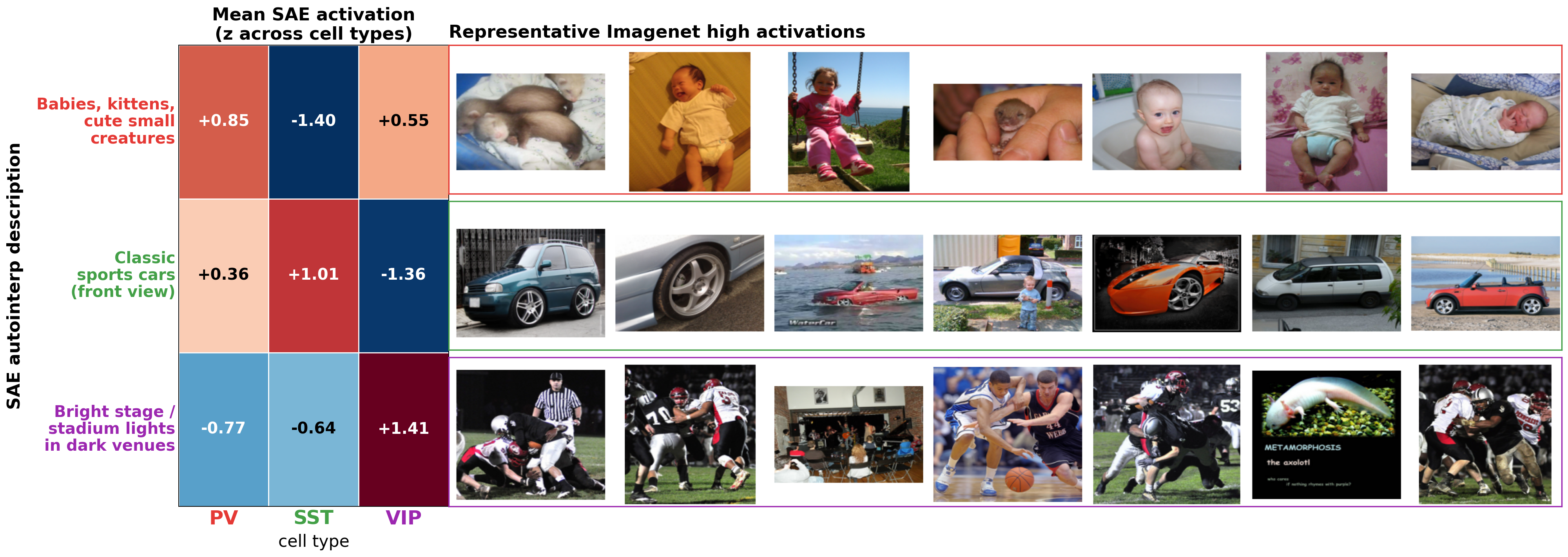}
    \caption{\textbf{Cell-type-unique SAE features map onto interpretable visual concepts.} Left: z-scored mean activation of five ``unique-by-magnitude'' SAE features across cell types. Right: top ImageNet-1k images per feature.}
    \label{fig:sae_dictionary_imagenet}
\end{figure*}

The resulting dictionary separates the three populations along interpretable axes (Fig.~\ref{fig:sae_dictionary_imagenet}). PV cells uniquely emphasise features for \emph{small rounded objects}: babies, kittens, teapots, toasters, household items. SST cells emphasise \emph{vehicles} --- specifically classic and sports cars, the dominant content of the NM1 noir movie. VIP cells emphasise something object-orthogonal: \emph{venue lighting and atmosphere}. Their unique features fire on produce displays under bright market light and on dark sports/concert arenas with stage illumination, with no consistent object category. The most distinctive VIP feature ($26984$, ``bright stage / stadium lights in dark venues'') activates $29\%$ more strongly in VIP than in PV or SST. Where SST differentiates the \emph{cars} in a scene, VIP differentiates the \emph{lighting} of that same scene. A lighting-and-atmosphere readout was not something we predicted \emph{a priori}: prior work establishes VIP interneurons as mediators of disinhibitory cortical gain control recruited by behavioral state and reinforcement signals \cite{fu2014cortical, pi2013cortical}, but does not specify what visual content their activity should covary with. Our finding is therefore best read as a tentative observation that may complement this literature (pointing to a possible link between VIP-mediated gain modulation and the encoding of scene-level luminance and contrast statistics) and one that requires direct circuit-level follow-up before a mechanistic claim is warranted.

\paragraph{Most features are shared; the small unique set is stable.} The picture from the dictionary is that PV, SST, and VIP \emph{share} most of what they encode (the bulk of their top-activating SAE features overlap heavily across cell types) but a small, population-specific tail of features is what actually sets them apart. We need to check that this tail is a real property of each population and not an artifact of the particular test bins we happened to evaluate on. The standard tool for this is a \emph{bootstrap}: we draw a new dataset of the same size as the original by sampling test bins \emph{with replacement} (so a given bin can appear several times in one bootstrap and not at all in the next), recompute everything from scratch on that resampled set, and repeat the procedure many times to see how stable the answer is. Concretely, we resample the $7{,}186$ NM1 test bins $200$ times, recompute mean activation per feature per cell-type on each resample, and re-rank the unique-by-magnitude features. The list barely changes: across all $200$ resamples, every cell-type recovers at least $15$ of its $20$ canonical features (mean overlap PV $18.4/20$, SST $17.7/20$, VIP $19.3/20$), and the $\sim$12 most specific features per cell-type appear in \emph{every single} resample (Fig.~\ref{fig:E6_bootstrap}, appendix). The same qualitative themes are also recovered from a probe that does not use ImageNet at all (a CLIP-text concept-axis probe; PV$\,\to\,$\textit{high contrast} $p{=}0.001$; SST$\,\to\,$\textit{vintage car} $p{=}0.002$; VIP$\,\to\,$\textit{shop window} $p{<}10^{-3}$; Appendix~\ref{app:concept_axis}).

Pointing \method at a labelled subset of neurons --- optotagged or otherwise --- therefore returns an interpretable concept dictionary in which each entry comes with both a quantitative weight and a set of natural-image exemplars, and in which the population-specific entries are stable under resampling. The PV / SST / VIP contrast above was produced end-to-end with no cell-type-aware training step: the encoder, the language decoder, and the SAE were all trained without ever being told which neuron belongs to which genetic line.

\section{Conclusion}
\label{sec:discussion}

We introduce \method, a single trained model that turns spike trains from arbitrary subsets of neurons in mouse visual cortex into free-form natural-language narrations of the viewed scene, with no language-side training and no stimulus-specific prior. The same model can be queried at the scale of thousands of neurons, of one cortical region, of a local population, or of a single molecularly-defined cell type, which makes the encoder, rather than the decoder, the unit of biological analysis. Beyond the capability itself, the framework introduces an evaluation pipeline that controls for memorization, temporal autocorrelation, and biological plausibility (held-out scenes, hippocampal control, shuffled labels, exclusion-zone retrieval), and a sparse-autoencoder analysis \cite{bricken2023monosemanticity, cunningham2023sparse, gao2024scaling, fel2025archetypal} that recovers interpretable concept-level features from the cell-type-specific decodings.

\paragraph{Limitations.} \method is trained on one species, on visual cortex, and on a small stimulus vocabulary; the cell-type analysis relies on optotagged populations of $40$--$100$ neurons pooled across mice via frame-aligned pseudo-mice, and we have not tested whether the cell-type contrasts hold within single animals. The decoded narrations describe gross scene content rather than fine perceptual detail, and our concept-axis validations are correlational. The sparse autoencoder is borrowed off-the-shelf from a CLIP-only pipeline and is not jointly trained with the neural encoder.

\paragraph{Future work.} The natural next steps are to extend \method beyond visual cortex and beyond mouse — to auditory and somatosensory recordings, to non-human-primate and human single-unit data, and to behaviour-rich paradigms where the decoded narration can be aligned with task variables and trial-by-trial choice. The greatest opportunity, however, may lie in modalities where humans lack an intuitive grasp of the stimulus space — most acutely olfaction, which has no agreed-upon parameterization of odor \cite{yeshurun2010odor}, and chemosensation more broadly. Precisely there, decoding neural activity directly into language may offer the shortest route to a representation humans can actually interpret. The cell-type interrogation pipeline extends to any labelled neural subset (genetic, anatomical, functional, or connectivity-defined), and coupling it with closed-loop optogenetic perturbation would turn the decoded concept dictionary into a causal handle. Finally, training the sparse bottleneck jointly with the neural encoder \cite{Qin2026SparseCC}, rather than reusing a pretrained CLIP SAE, opens the door to discovering concepts present in neural activity but absent from the vision-language prior — the path we view as most promising toward using language models as systematic instruments for neuroscience discovery.

\section{Acknowledgements}
This work was funded by Harvard Mind, Brain, Behavior Interfaculty Initiative (\url{https:// mbb.harvard.edu/}). Arnau Marin-Llobet is supported by Coefficient Giving and the RCC-Harvard Fellowship. Richard Hakim is supported by Kempner Research Fellowship. 

\FloatBarrier
{
    \small
    \bibliography{main.bib}
    \bibliographystyle{unsrt}
}
\FloatBarrier

\appendix
\counterwithin{figure}{section}
\counterwithin{table}{section}
\renewcommand{\thefigure}{A\arabic{figure}}
\renewcommand{\thetable}{A\arabic{table}}
\clearpage
\section{Appendix}

This appendix collects supporting material referenced from the main text. Section~\ref{app:method_details} provides full architectural and training-configuration details for \method. Section~\ref{app:datasets} provides full dataset and experimental-procedure details. Section~\ref{app:narration_examples} presents extended frame-by-frame narration examples from a second movie (NM3). Section~\ref{app:held_out} provides additional held-out evaluation detail. Section~\ref{app:benchmarks} reports baseline comparisons and embedding-quality controls. Section~\ref{app:celltype_extra} extends the concept-axis analyses for the cell-type populations. Section~\ref{app:sae_robustness} documents the bootstrap-stability and concept-axis controls used to validate the cell-type SAE feature dictionary.

\subsection{Extended narration examples}
\label{app:narration_examples}
The two figures below show frame-by-frame decoded narrations from \method on a longer NM1 segment and on the unrelated movie clip NM3. Same conventions throughout: top row shows movie frames, middle row shows representative single-unit spike rasters from the corresponding session, bottom row shows the decoded narration generated from neural activity alone (no language-side training, no stimulus prior).

\begin{figure*}[ht]
    \centering
    \includegraphics[width=\linewidth]{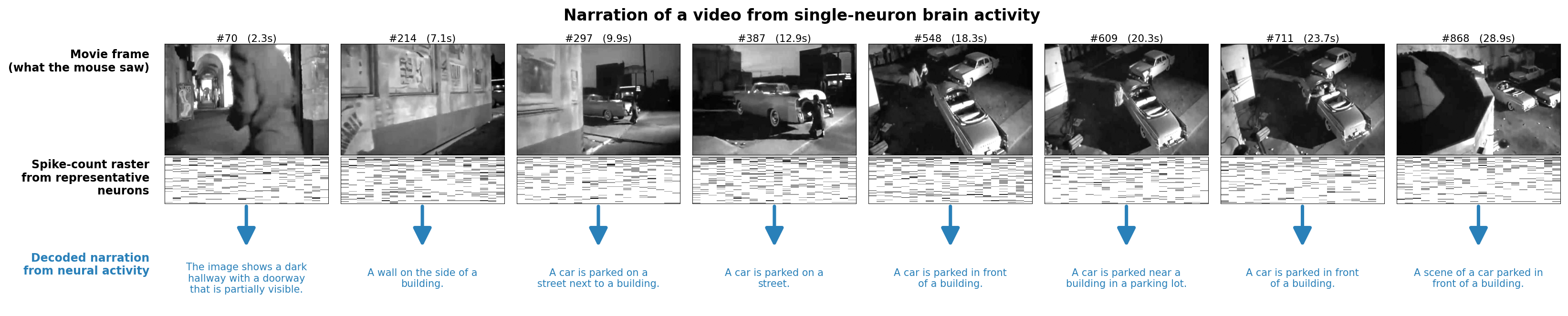}
    \caption{\textbf{Frame-by-frame narration on NM1.} Top: movie frames.
    Middle: spike rasters. Bottom: \method narrations from neural activity
    alone.}
    \label{fig:video_narration}
\end{figure*}

\begin{figure*}[ht]
    \centering
    \includegraphics[width=\linewidth]{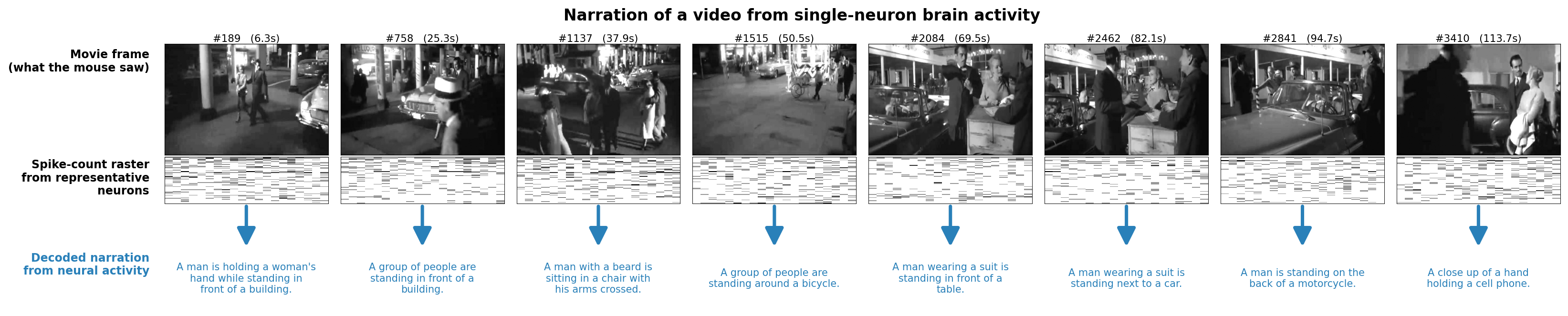}
    \caption{\textbf{Frame-by-frame narration on NM3.} Same conventions
    as Fig.~\ref{fig:video_narration}.}
    \label{fig:video_narration_nm3}
\end{figure*}

\subsection{Method implementation details}
\label{app:method_details}

\paragraph{Spike preprocessing.}
Spike times from all single units passing Allen-Institute quality control are binned at {120}{Hz} and $z$-scored per neuron using mean and standard deviation computed from the training repeats only (validation and test repeats are excluded from preprocessing statistics). At inference, a {167}{ms} window of binned activity ($\approx\!20$ bins) is fed to the encoder.

\paragraph{Encoder architecture.}
The {\method Encoder} is a $12.8$\,M-parameter network with four stages:
(i) a multi-scale 1-D-convolutional front end with three parallel branches (kernel sizes $\{3, 7, 15\}$, $128$ channels each), capturing spike features at fast, intermediate, and slow time-scales;
(ii) a $2$-layer transformer encoder ($d_\text{model}\!=\!384$, $8$ heads) integrating information across the temporal window;
(iii) attention-weighted pooling collapsing the time axis to a single context vector;
(iv) $576$ learned patch queries that cross-attend to the pooled representation to produce one $1024$-dimensional embedding per CLIP patch.
Dropout is applied at $20\%$ throughout, and all nonlinearities are GELU.

\paragraph{Patch-tensor shape.}
The output shape ($576$ tokens of dimension $1024$) is fixed by CLIP ViT-L/14: the vision tower tiles a $336\!\times\!336$ image into $14$-pixel patches, yielding a $24\!\times\!24$ grid of patch tokens, and $1024$ is the vision encoder's hidden dimension at the penultimate layer.

\paragraph{Optimization.}
We optimize with AdamW (\textsc{lr} $=3\!\times\!10^{-4}$, weight decay $10^{-3}$), a cosine learning-rate schedule, and gradient clipping at $1.0$. Batch size is $64$ on the patch loss. Training runs for up to $60$ epochs with $12$-epoch early stopping on a held-out split of the training repeats. Training was performed on a single NVIDIA A100 (40GB) for ~1 hours per run.

\paragraph{Inference configuration.}
LLaVA-1.5-7B runs in its default greedy-decoding configuration with a maximum of $60$ output tokens and the fixed prompt
\texttt{"USER: <image>\textbackslash n Describe this scene in one sentence.\textbackslash n ASSISTANT:"}.
The PatchInjector forward hook overwrites the vision-tower output tensor in place; no changes are made to the multimodal projector or the LLaMA-2-7B decoder, both of which run in their released configuration.

\subsection{Dataset and experimental procedures}
\label{app:datasets}

\paragraph{Source dataset and session selection.}
All neural data come from the Allen Brain Observatory Visual Coding Neuropixels release \cite{siegle2021allen}, restricted to the Brain Observatory 1.1 (BO~1.1) subset --- the only stimulus protocol in the release that includes naturalistic stimuli. We use $16$ recording sessions, selected for completeness of the naturalistic-stimulus blocks and for the presence of at least one optotagged neuron of interest in the cell-type analyses. Stimulus-aligned spike counts are precomputed per session and split by stimulus repeat into train / validation / test partitions; the same partition is reused across all analyses in the paper.

\paragraph{Stimulus classes.}
Each mouse views three classes of natural content during the same recording session, allowing within-animal comparison of movie- and image-driven responses:
\textit{Natural Movie One} (NM1) is a {30}{s} clip presented at {30}{Hz} ($900$ unique frames, $20$ repeats per session);
\textit{Natural Movie Three} (NM3) is a {120}{s} clip ($3{,}600$ unique frames at {30}{Hz}, $10$ repeats);
\textit{Natural Scenes} is a fixed bank of $118$ distinct grayscale photographs, each presented for {250}{ms} interleaved with gray-screen trials, with $m\!50$ presentations per image. NM1 and NM3 share the {30}{Hz} frame cadence but differ in clip length, content, and repeat structure. The $50$ held-out trials reported in Appendix~\ref{app:image_holdout} are drawn from the $18$ Natural-Scenes images held out under the identity-holdout split (seed~$42$, fixed across sessions).

\paragraph{Optotagging and cell-type criteria.}
For all cell-type analyses (Sec.~\ref{sec:sae_dict}, Appendix~\ref{app:celltype_extra}, Appendix~\ref{app:sae_robustness}) we intersect the BO~1.1 sessions with the optotagging tables released by Siegle et al.\ \cite{siegle2021allen} and retain only neurons that pass their standard criteria: response reliability $>30\%$ to the blue-light pulse, median first-spike latency $<\!{8}{ms}$, and response rate at least $2\times$ baseline. Across the $16$ sessions this yields $73$ parvalbumin (PV) cells from $5$ \textit{Pvalb-IRES-Cre} mice, $49$ somatostatin (SST) cells from $6$ \textit{Sst-IRES-Cre} mice, and $33$ vasoactive-intestinal-peptide (VIP) cells from $5$ \textit{Vip-IRES-Cre} mice. Because each transgenic line targets a single interneuron class, no animal contributes neurons of more than one cell type, and the three populations are fully disjoint at the animal level.

\paragraph{Pseudo-mouse construction.}
Optotagged populations within a single session are too small to support population-level decoding (median $m=7$ cells per session for the rarest class). We therefore construct frame-aligned \emph{pseudo-mice} by concatenating the optotagged columns of the spike-count matrix across all sessions of matching genotype. This is well-defined because the BO~1.1 protocol presents identical stimulus sequences to every mouse: empirical frame misalignment between sessions is $<0.1\%$, well below the {33}{ms} frame bin. The resulting pseudo-mouse has one row per stimulus frame and one column per optotagged neuron pooled across animals of the same line, and is treated as a single recording for all downstream training and evaluation.

\subsection{Held-out semantic alignment}
\label{app:held_out}
\begin{figure}[t]
    \centering
    \includegraphics[width=\linewidth]{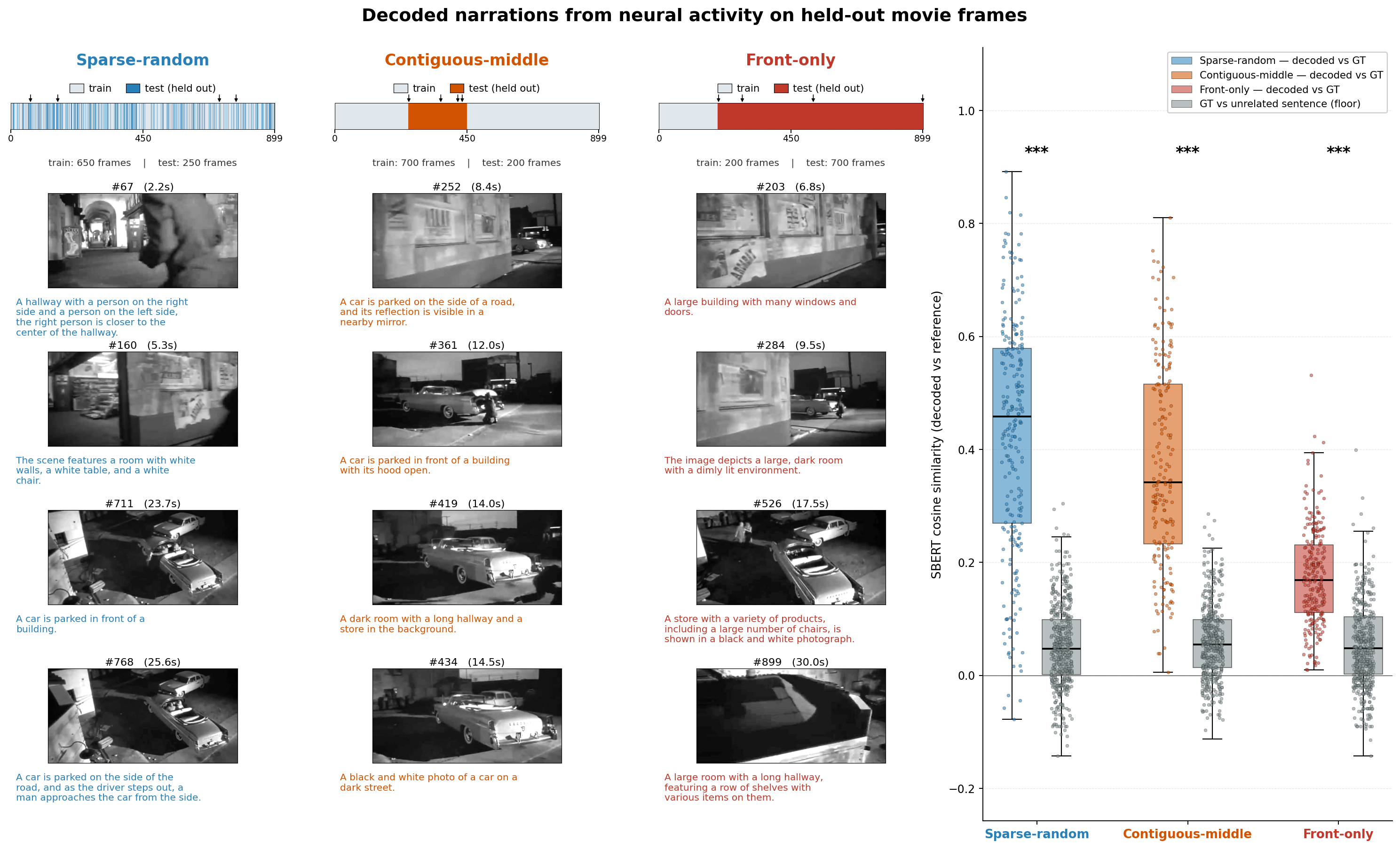}
    \caption{\textbf{Decoded narrations are semantically aligned with
    the true frame.} Per-frame SBERT cosine to the reference caption
    (colored) vs random-sentence null (gray).}
    \label{fig:holdout_sbert}
\end{figure}

Fig.~\ref{fig:holdout_sbert} reports per-frame SBERT cosine between \method's decoded narrations and the BLIP-2 reference caption for the same frame, alongside a random-sentence floor obtained by pairing each reference with topic-unrelated sentences from a fixed pool. The boxplot complements Fig.~\ref{fig:held_out_narration_quality} in the main text by showing the per-frame distribution rather than aggregate means.

\subsection{Generalization to never-seen images} \label{app:image_holdout} 
\begin{figure}[t]
    \centering
    \includegraphics[width=\linewidth]{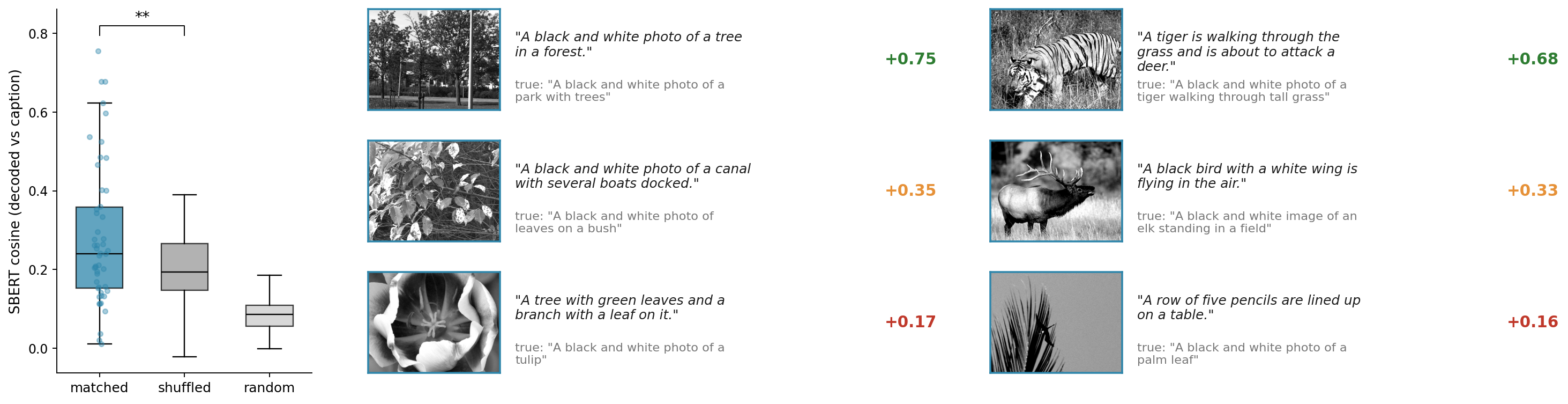}
    \caption{\textbf{Decoding novel image identities.} \method is tested on $18$ Allen Natural Scenes images held out entirely from training (responses seen only at test time). \textbf{Left:} SBERT cosine between decoded narration and the true BLIP-2 caption, for matched pairings, shuffled pairings (decoded sentence vs.\ a random other held-out caption), and a random-sentence floor. Matched decodes score significantly higher than shuffled (Wilcoxon signed-rank, $p\!<\!10^{-2}$). \textbf{Right:} example trials on unseen images, ordered from accurate (top) to failed (bottom). Decoded narrations (italic) are shown above the true captions (grey), with SBERT cosine on the right. Successful decodes recover scene content (tree-in-forest, tiger-in-grass); near-misses capture the right semantic neighbourhood but wrong specifics (elk read as a bird); failures land on unrelated content (palm leaf decoded as pencils).}
    \label{fig:image_id_retrieval}
\end{figure}

Held-out frames within a familiar movie share statistics with their training neighbours; a stronger test of generalization is whether \method can decode neural responses to \emph{novel image identities} that the encoder has never been exposed to. We use the Allen Natural Scenes stimulus, a fixed bank of $118$ unrelated grayscale photographs. Following an identity-holdout split (seed~$42$, fixed across sessions), the encoder is trained on responses to $100$ of these images and $18$ images are held out entirely --- their spike responses appear \emph{only} at test time. At evaluation, decoded narrations are scored via SBERT against the BLIP-2 reference caption of the true held-out image (Fig.~\ref{fig:image_id_retrieval}). 

On $50$ held-out trials, decoded narrations achieve $\mathrm{SBERT}=0.282 \pm 0.178$ against the true caption, versus $0.222 \pm 0.136$ for a shuffled-pairing control (decoded sentences paired with a random other held-out caption) and $0.087 \pm 0.049$ for a random-sentence floor (Wilcoxon signed-rank, matched~vs.~shuffled $p\!<\!10^{-2}$; $\Delta_{\text{matched}-\text{shuffled}}\!=\!+0.060$, $\Delta_{\text{matched}-\text{floor}}\!=\!+0.195$). $45$ of $50$ decoded narrations are distinct sentences --- the diversity expected from genuine per-trial decoding rather than collapse onto a single default response. Inspection of individual trials (Fig.~\ref{fig:image_id_retrieval}, right) shows content-accurate decodes of unseen images: a park-with-trees scene narrated as \textit{``a tree in a forest''} ($+0.75$), a tiger-in-tall-grass scene as \textit{``a tiger walking through the grass''} ($+0.68$), and a stairs scene as \textit{``a spiral staircase with a metal railing''} alongside near-misses where the decoded narration captures the right semantic neighbourhood but the wrong specifics (an elk read as \textit{``a black bird with a white wing flying in the air''}, $+0.33$). These results indicate that the spike-to-language mapping carries information about the visual content of stimuli the encoder was never trained on, rather than reflecting memorisation of the training images.

\subsection{Baseline comparisons}
\label{app:benchmarks}
We compare \method against three families of baselines on the same NM1 test frames used in the main text. For the retrieval-only entries, the patch head is replaced by a single $512$-d projection trained with a $70/30$ mixture of cosine similarity and InfoNCE loss against CLIP-text embeddings of the same frames; everything else (encoder, splits, optimizer) is held fixed. Tab.~\ref{tab:decoding_quality} reports decoding quality across three holdout regimes, Tab.~\ref{tab:embedding_quality} reports embedding-level retrieval quality (CKA, KNN purity, R@10, median rank), and Tab.~\ref{tab:retrieval_benchmark} reports frame-level R@1 across four pilot sessions for an additional CEBRA \cite{schneider2023learnable} dimensionality sweep.

\begin{table}[h]
\centering
\caption{\textbf{Decoding quality across architectures and training schemes.} Per-frame SBERT cosine between decoded sentences and BLIP-2 captions on held-out NM1 frames (mean $\pm$ SD). Three holdout regimes: sparse random frames, a contiguous middle block, and front-only extrapolation. Random-sentence floor: SBERT cosine of each decoded caption against a fixed pool of 30 off-topic sentences.}
\label{tab:decoding_quality}
\small
\begin{tabular}{lcccc}
\toprule
Method & Sparse random & Contiguous middle & Front-only & Mean \\
\midrule
\textbf{\method (ours)}        & \textbf{0.426 $\pm$ 0.21} & \textbf{0.367 $\pm$ 0.18} & \textbf{0.170 $\pm$ 0.09} & \textbf{0.32} \\
CEBRA + LLaVA                  & 0.063 $\pm$ 0.05 & 0.069 $\pm$ 0.05 & 0.069 $\pm$ 0.05 & 0.07 \\
Ridge + LLaVA                  & 0.047 $\pm$ 0.05 & 0.061 $\pm$ 0.05 & 0.076 $\pm$ 0.05 & 0.06 \\
Random-sentence floor          & 0.007 $\pm$ 0.08 & 0.020 $\pm$ 0.08 & 0.062 $\pm$ 0.07 & 0.03 \\
\bottomrule
\end{tabular}
\end{table}

\begin{table}[h]
\centering
\caption{\textbf{Embedding retrieval quality} on held-out NM1 frames (sparse-random condition, 900 frames). CKA: linear Centered Kernel Alignment between predicted features and true CLIP ViT-L/14 pooled features (in $[0, 1]$). KNN purity: fraction of the $K{=}10$ nearest test bins (in predicted feature space) that share the query's true frame (chance $\approx 1/900$). R@10 / median rank: cosine retrieval of the true frame in CLIP-L pooled space (chance R@10 $\approx 1.1\%$, chance median rank $= 450$); only defined when the predicted feature lives in CLIP-L's $1024$-d patch space.}
\label{tab:embedding_quality}
\small
\begin{tabular}{lcccc}
\toprule
Method & CKA $\uparrow$ & KNN purity ($K{=}10$) $\uparrow$ & R@10 $\uparrow$ & Median rank $\downarrow$ \\
\midrule
\textbf{\method (ours)}     & \textbf{0.890} & \textbf{0.174} & \textbf{38.8\%} & \textbf{14} \\
CEBRA (w/ our training)     & 0.016 & 0.030 & 1.3\%  & 436 \\
CEBRA (naive)               & 0.094 & 0.147 & --- & --- \\
MLP-Patches                 & 0.758 & 0.162 & 21.0\% & 35 \\
\bottomrule
\end{tabular}
\end{table}

\begin{table}[h]
\centering
\caption{\textbf{Frame-level R@1 retrieval on NM1 across four pilot
sessions} (chance $= 1/900 = 0.11\%$). Methods compared: ridge
regression on raw spike windows, PCA followed by KNN retrieval, and
CEBRA at three output dimensions decoded with KNN.}
\label{tab:retrieval_benchmark}
\small
\begin{tabular}{lccccccc}
\toprule
Method & Params & Sess 3703 & Sess 3822 & Sess 8571 & Sess 8357 & Mean \\
\midrule
Ridge                          & 1.08M  & 1.3\%  & 0.8\%  & 0.5\%  & 0.4\%  & 0.7\%  \\
PCA+KNN                        & 0.27M  & 1.6\%  & 1.0\%  & 0.3\%  & 0.4\%  & 0.8\%  \\
CEBRA-128 + KNN                & 0.74M  & 8.5\%  & 6.0\%  & 4.5\%  & 2.0\%  & 5.2\%  \\
CEBRA-256 + KNN                & 0.81M  & 12.0\% & 9.8\%  & 5.5\%  & 3.1\%  & 7.6\%  \\
CEBRA-512 + KNN                & 0.89M  & 16.1\% & 12.9\% & 7.8\%  & 4.3\%  & 10.3\% \\
Chance                         & ---    & 0.11\% & 0.11\% & 0.11\% & 0.11\% & 0.11\% \\
\bottomrule
\end{tabular}
\end{table}

\subsection{Additional cell-type concept-similarity curves}
\label{app:celltype_extra}
\begin{figure*}[t]
    \centering
    \includegraphics[width=\linewidth]{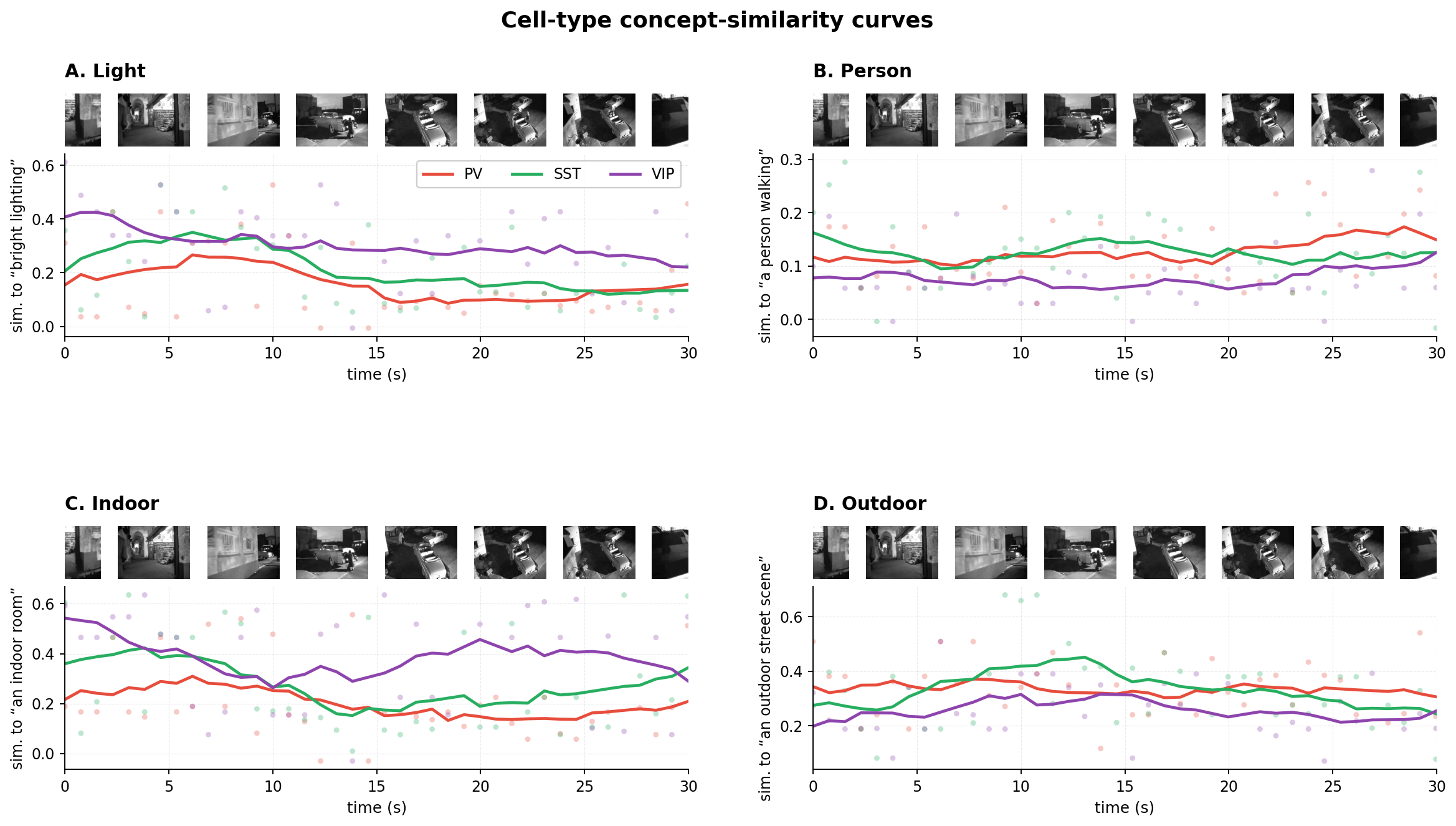}
    \caption{\textbf{Additional concept-similarity curves per cell
    type.} Time-resolved cosine of PV/SST/VIP narrations to (A)
    \textit{``bright lighting''}, (B) \textit{``a person walking''},
    (C) \textit{``an indoor room''}, (D) \textit{``an outdoor street
    scene''}. Same conventions as Fig.~\ref{fig:celltype_main}B--C.}
    \label{fig:celltype_supp}
\end{figure*}

Fig.~\ref{fig:celltype_supp} extends Fig.~\ref{fig:celltype_main}B--C to four additional concept axes (\textit{``bright lighting''}, \textit{``a person walking''}, \textit{``an indoor room''}, \textit{``an outdoor street scene''}). PV, SST, and VIP narrations are projected onto each text-concept embedding using CLIP-text. The qualitative ordering of cell types reproduces across concept axes within visual modality.

\subsection{Robustness of the cell-type SAE feature dictionary}
\label{app:sae_robustness}
We perform two controls on the cell-type-unique SAE feature dictionary reported in Sec.~\ref{sec:sae_dict}: a bootstrap-stability test that verifies the dictionary is not driven by outlier test bins, and a CLIP-text concept-axis probe that recovers the dictionary's qualitative themes from an analysis pipeline that does not look at narration content.

\begin{figure*}[t]
    \centering
    \includegraphics[width=\linewidth]{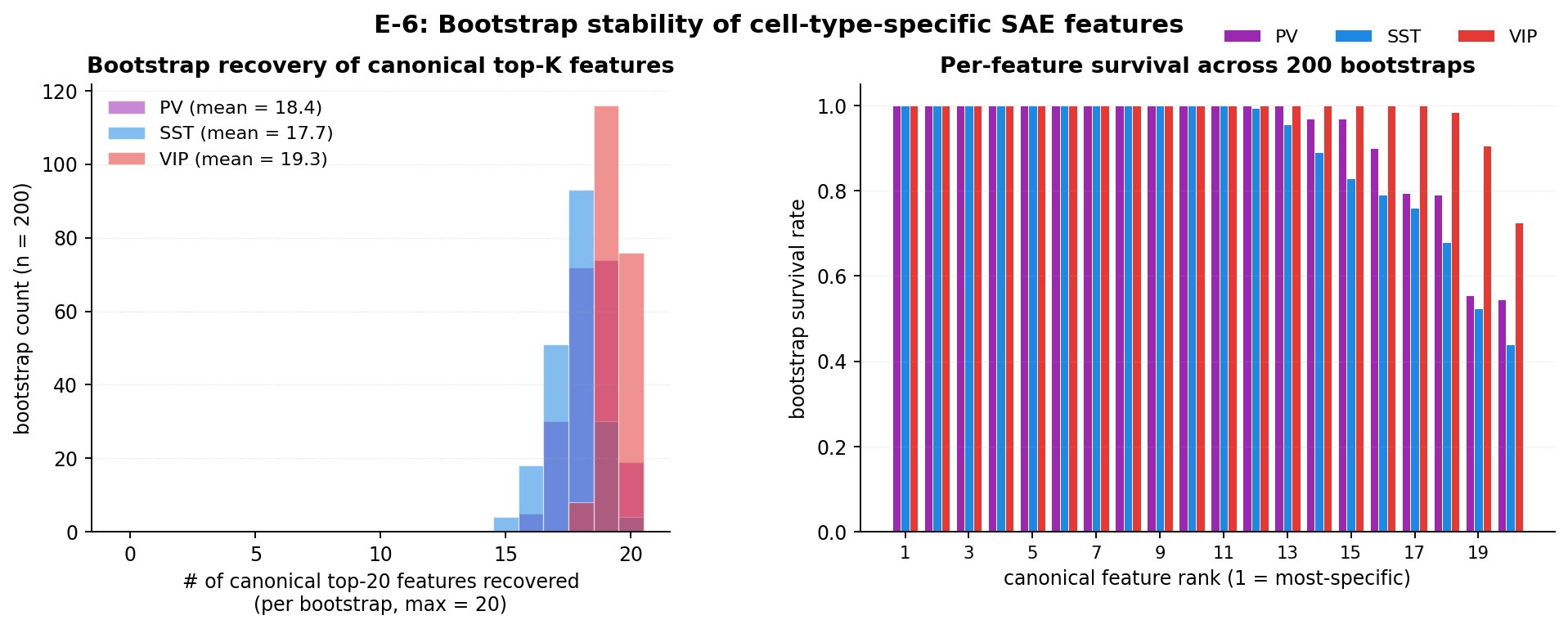}
    \caption{\textbf{Cell-type-unique SAE features are stable to bin
    resampling.} Left: number of canonical top-20 features recovered
    per bootstrap (max 20; $n{=}200$). Right: per-feature survival
    rate, sorted by canonical specificity rank.}
    \label{fig:E6_bootstrap}
\end{figure*}

\paragraph{Bootstrap stability.}
\label{app:bootstrap_stability}
We resample the $7{,}186$ NM1 test bins with replacement
($n{=}200$ bootstraps), recompute mean activation per feature per
cell type, and re-rank specificity from scratch each time. For every cell type, the bootstrapped top-20 recovers at least $15$ of the canonical $20$ features in $100\%$ of resamples (mean overlap: PV $18.4/20$, SST $17.7/20$, VIP $19.3/20$). The most-specific $m$12 features per cell type survive every resample
(Fig.~\ref{fig:E6_bootstrap}).

\begin{figure*}[t]
    \centering
    \includegraphics[width=\linewidth]{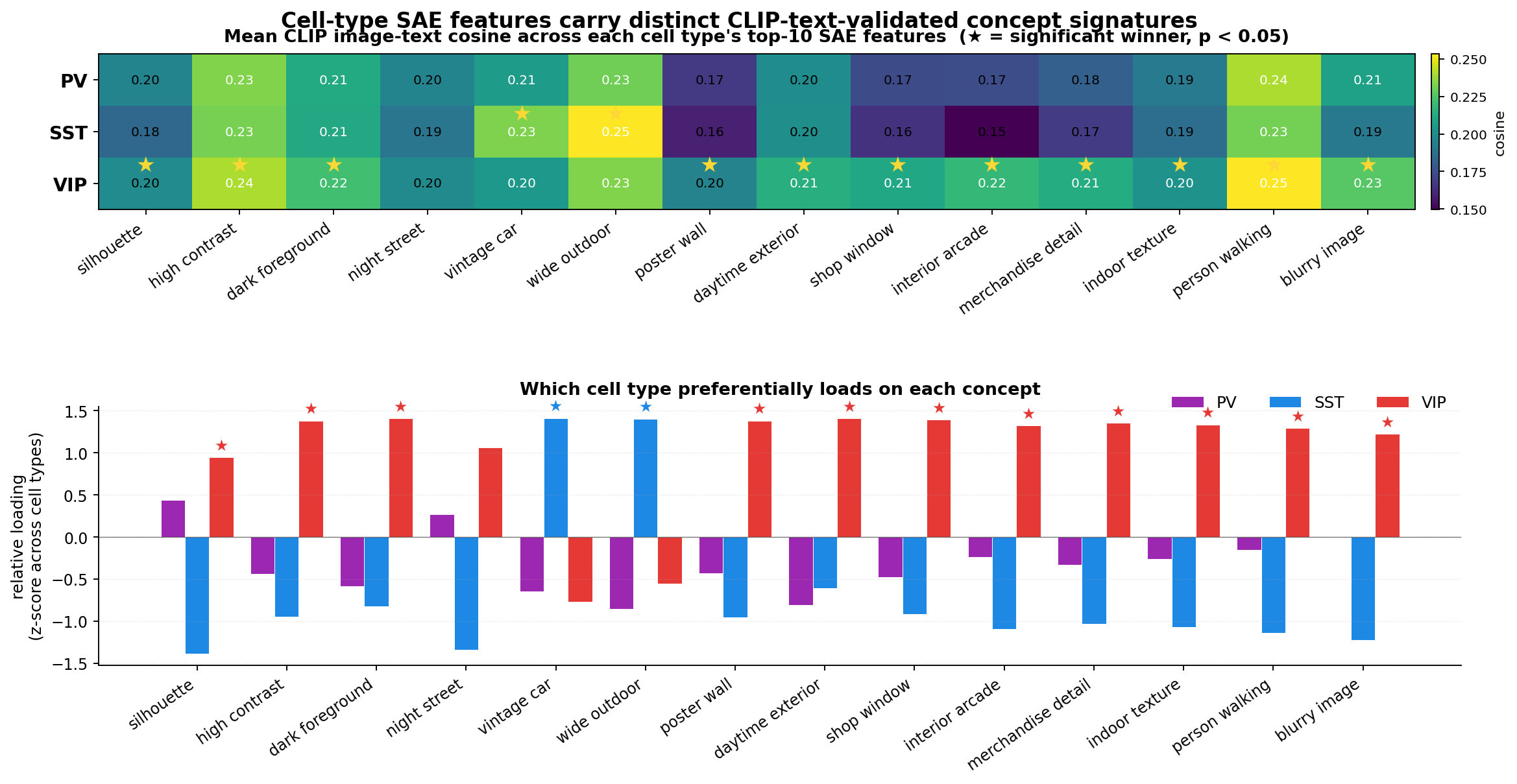}
    \caption{\textbf{CLIP-text concept-axis validation of cell-type
    SAE features.} Top: mean CLIP image-text cosine between each
    cell type's top-10 SAE features and 14 concept prompts. Bottom:
    same loadings z-scored across cell types. Stars: Welch's
    $t$-test, $p<0.05$, one-vs-rest.}
    \label{fig:sae_concept_validation}
\end{figure*}

\paragraph{CLIP-text concept-axis validation.}
\label{app:concept_axis}
As a probe orthogonal to the ImageNet auto-interpretation in Sec.~\ref{sec:sae_dict}, we measure the mean CLIP image-text cosine between each cell type's top-10 SAE features (each represented by its top-5 maximally-activating NM1 frames) and a curated bank of $14$ visual concepts (silhouette, high contrast, vintage car, shop window, interior arcade, etc.). The relative loading recovers the qualitative themes of Sec.~\ref{sec:sae_dict}: the top-loading concept per cell type is significantly higher than the same concept's loading on the other cell types' features (Welch's $t$-test, PV$\,\to\,$\textit{high contrast} $p=0.001$; SST$\,\to\,$\textit{vintage car} $p=0.002$; VIP$\,\to\,$\textit{shop window} $p<10^{-3}$; Fig.~\ref{fig:sae_concept_validation}).

\newpage

\end{document}